\documentclass[12pt]{iopart}

\usepackage{graphics}
\usepackage{color}
\usepackage{psfrag}
\usepackage{amssymb}
\usepackage{subfigure}

\begin{document}

\title{ Retrieving information from a noisy 
``knowledge network''}

\author{J. Barr\'e}

\address{Universit\'e de Nice-Sophia Antipolis, Laboratoire J.~A.~Dieudonn\'e 
UMR CNRS 6621, Parc Valrose 06108 Nice Cedex 02, France.}
\ead{jbarre@unice.fr}

\begin{abstract}
  We address the problem of retrieving information from a noisy version
  of the ``knowledge networks'' introduced by Maslov and
  Zhang~\cite{zhang}. We map this problem onto a disordered
  statistical mechanics model, which opens the door to many analytical
  and numerical approaches. We give the replica symmetric solution,
  compare with numerical simulations, and finally discuss an
  application to real data from the United States Senate.
\end{abstract}
%Uncomment for PACS numbers title message
%\pacs{}
% Keywords required only for MST, PB, PMB, PM, JOA, JOB? 
\vspace{2pc}
\noindent{\it Keywords}: Communication, supply and information networks; 
Random graphs, networks; Message-passing algorithms.
% Uncomment for Submitted to journal title message
%\submitto{\JPA}
% Comment out if separate title page not required
\maketitle

\section{Introduction}

In a recent paper~\cite{zhang}, Maslov and Zhang addressed the
following problem: we are given $N$ agents, each one represented by an
$M$-dimensional real vector $\vec{r}_i$; suppose we know $K$ of the
$N(N-1)/2$ scalar products $\Omega_{ij}=\vec{r}_i.\vec{r}_j$ with
$i \neq j$. In this situation, can we predict the value of an unknown
scalar product $\Omega_{ij}$?  This question is relevant for instance
to the problem of extracting information from the vast amount of data
generated by a commercial website. The $\vec{r}_i$ may represent in
that context the interests of a person $i$, and $\Omega_{ij}$ the
mutual appreciation of persons $i$ and $j$; the problem is then to
predict the mutual appreciation of two persons that do not know each
other. Maslov and Zhang called the network of interactions and
overlaps $\Omega_{ij}$ a ``knowledge network''\footnote{These authors
  actually introduce a bipartite version of these networks.}.

One of their main results is the following: there exists a critical
density of known overlaps $p_c=2K/N(N-1)$ above which almost all the a
priori unknown overlaps are completely determined by the $K$ known
ones. This transition is a realization of the so-called rigidity
percolation. However, their treatment leaves several important issues
aside, and assumes that we have at our disposal much more information
that we typically do. For instance, the size of the vectors $M$
describing each agent is a priori unknown; the problem of estimating
$M$ from the data was addressed in~\cite{franco}. More drastically,
the data on the overlaps is necessarily noisy: if $\vec{r}_i$ and
$\vec{r}_j$ model the interests of persons $i$ and $j$, their mutual
appreciation $\Omega_{ij}$ is certainly not completely determined by
the overlap of their interests $\vec{r}_i.\vec{r}_j$, although it is
probably biased by it. In this more realistic case of noisy
information, the questions are: does the ``phase transition'' noted by
Maslov and Zhang survive? And how to retrieve the information
contained in the noisy knowledge network?  We address these issues in
the following by studying a simple model of this situation.

%By exhibiting a mapping of this situation onto a disordered
%statistical mechanics problem, we address these issues in the
%following.

The outline of the paper is as follows: we present in
section~\ref{sec:model} the details of the model we are going to
study, and the mapping onto a disordered statistical mechanics
problem, which happens to be the one studied in~\cite{Thouless} and
more recently in~\cite{castellani}. This mapping opens the door to the
use of many analytical and numerical methods. In
section~\ref{sec:cavity}, we give the solution of this problem at the
replica symmetric level, using the cavity method~\cite{cavity}. We
then check these analytical results against numerical simulations in
section~\ref{sec:num}, and real data from the United States Senate in
section~\ref{sec:US_senate}.

%Customers preferences; vast amount of datas; exploitation?\\ 
%Maslov-Zhang: knowledge network; summary of the results.\\
%question: no noise! They themselves recognize that in any real life 
%situation, some noise would be present.\\
%What happens when information is not perfect? Still a phase transition?
%How to retrieve the information?\\
%Outline: presentation of the model; cavity solution; numerical 
%simulations; application on a real data set; conclusion.

\section{The model}
\label{sec:model}

We present now the noisy version of Maslov and Zhang's ``knowledge
network'' which we are going to study; for simplicity, the variable
describing each agent is discrete, and one dimensional. We consider $N$
agents; each one is characterized by an opinion $s_i^0$, with
$i=1,\ldots,N$; the $s_i^0$ may take $k$ different values, and are a
priori unknown. The $s_i^0$ may be for instance political opinions, as
in the example of section~\ref{sec:US_senate}.  We suppose we have
some information on the $s_i^0$, given by a an analog of the
``overlaps'' of~\cite{zhang}~: for a certain number of pairs $(i,j)$
we know a number $J_{ij}$ associated to it, constructed as follows.
If $s_i^0=s_j^0$, then $J_{ij}=1$ with probability $1-p$, and
$J_{ij}=-1$ with probability $p$; if $s_i^0\neq s_j^0$, then $J_{ij}=1$
with probability $p$, and $J_{ij}=-1$ with probability $1-p$. We take
$p\leq 1/2$. $p$ is then a measure of the noise in the information; in
the limit $p=1/2$, the network does not convey any information on the
$s_i^0$. The basic questions we ask are: how well can we reconstruct
the actual opinions $s_i^0$ knowing the $J_{ij}$? Do we have an
effective algorithm to do so?

We are interested in the probability of any set of opinions
$\{s_i\}_{i=1,\ldots,N}$, given the $J_{ij}$ representing our knowledge;
from Bayes formula, we can write:
\begin{equation}
P \left( \{s_i\}|\{J_{ij}\} \right) = \frac{P(\{s_i\})}{P(\{J_{ij}\})}
P \left( \{J_{ij}\}| \{s_i\}\right)~.
\label{eq:proba}
\end{equation}
  
The factor $P(\{s_i\})$ is the prior probability on the $s_i$; we
suppose from now on that it is flat, so that this term is independent
of the $s_i$. It would be possible however to consider another prior
probability. The factor $P(\{J_{ij}\})$ is difficult to compute, as the
$J_{ij}$ are correlated in an intricated way; however, it is in any
case independent of the $s_i$, so it acts as a normalization factor
for the distribution~(\ref{eq:proba}). Finally, the $P \left(\{J_{ij}\}|
  \{s_i\}\right)$ is easy to compute, since once the $s_i$ are given,
the $J_{ij}$ are independent. Let us consider two agents $1$ and $2$
with opinions $s_1$ and $s_2$; then from simple algebra one checks
that
\begin{equation}
P(J_{12}|s_1,s_2)=\sqrt{\frac{1-p}{p}} \left( \sqrt{\frac{1-p}{p}}
\right)^{\frac{1}{2}J_{12}(2\delta_{s_1,s_2}-1)}~.
\label{eq:probaJ}
\end{equation}
Since the $J_{ij}$ are independent once the $s_i$ are given,
Eq.~(\ref{eq:proba}) may be rewritten as
\begin{equation}
P \left( \{s_i\}|\{J_{ij}\} \right) \propto \Pi_{<i,j>} \left( \sqrt{\frac{1-p}{p}}
\right)^{\frac{1}{2}J_{ij}(2\delta_{s_i,s_j}-1)}~,
\end{equation}
where the index $<i,j>$ means that the product runs over the pairs
$(i,j)$ that are connected by a known $J_{ij}$. 
Taking the logarithm, we have:
\begin{equation}
H[\{s_i\}]=-Log \left[ P \left( \{s_i\}|\{J_{ij}\} \right) \right] = 
\mbox{Cste}-B\sum_{<i,j>} J_{ij}(2\delta_{s_i,s_j}-1)~,
\label{eq:ham}
\end{equation}
with
$$
B=\frac{1}{2}Log\left( \frac{1-p}{p}\right)~.
$$

Eq.~(\ref{eq:ham}) can be seen as the Hamiltonian of a disordered
Potts model, which opens the door to the use of many analytical and
numerical tools to study it. From now on, we will concentrate for
simplicity on the Ising case, where each agent may have only two
opinions, $s_i=+1$ or $s_i=-1$. In this Ising case, the Hamiltonian
reads:
\begin{equation}
H[\{s_i\}]=-Log \left[ P \left( \{s_i\}|\{J_{ij}\} \right) \right] = 
\mbox{Cste}-B\sum_{<i,j>} J_{ij}s_is_j~,
\label{eq:ham_is}
\end{equation}
The sets $\{s_i\}$ with maximum probability are the minimizers of
Eq.~(\ref{eq:ham}); the minimizer is not necessarily unique.  The
question, how well can we reconstruct the real opinions knowing the
$J_{ij}$ is then rephrased as: given a minimizer $\{s_i^{\ast}\}$ of
Eq.~(\ref{eq:ham}), how far is it from the real opinions $\{s_i^0\}$?
We answer this question in the next section.  We note that this
rephrasing of the problem bears some resemblance with the community
detection, or clustering problem as stated in~\cite{hastings}; in this
work however, the probabilistic analysis yields a Potts-like model
without disorder.

\section{Cavity solution}
\label{sec:cavity}

\subsection{Gauge transformation}

Hamiltonian~(\ref{eq:ham_is}) is not as well-suited for analytical
treatment as it seems to be. It is a disordered Ising model, but the
probability distribution of the couplings $J_{ij}$ is not known, and
actually very complicated: the relevant information we want to extract
is precisely hidden in the correlations between the $J_{ij}$. 
The following gauge transformation, somewhat miraculously, yields a 
tractable problem.

We define $\tilde{s}_i=s_i^0s_i$ and $\tilde{J}_{ij}=s_i^0s_j^0J_{ij}$
(the $s_i^0$ are the true opinions of the agents); then
\begin{equation}
H[\{s_i\}]= -B\sum_{<i,j>} \tilde{J}_{ij}\tilde{s}_i\tilde{s}_j~.
\label{eq:ham2}
\end{equation}
The distribution of the $\tilde{J}_{ij}$ does not depend any more on
the ${s_i^0}$: $\tilde{J}_{ij}=1$ with probability $1-p$ and
$\tilde{J}_{ij}=-1$ with probability $p$: all correlations in the
couplings have disappeared. Furthermore, given a set $\{\tilde{s}_i\}$, it
is easy to know how far the corresponding set $\{s_i\}$ is from the
original $\{s_i^0\}$: it is enough to compute the number of
$\tilde{s}_i$ equal to $-1$. Thus we are left with the study of
Hamiltonian~(\ref{eq:ham2}), which is that of a ferromagnetically
biased Ising spin glass. We would like to compute the magnetization of
the ground state of such a Hamiltonian. From now on, we remove the
$\sim$ on the $J$'s and the $s$'s. Let us note that the ground state
does not depend on $B$, so we may take $B=1$ for simplicity (as long
as $B>0$, that is $p<1/2$). All explicit dependence on $p$ is then
removed, which is very convenient for practical purposes, as $p$ is a
priori unknown: the knowledge of the $J_{ij}$ is sufficient to
determine the minimizers of~(\ref{eq:ham2}). We need however to keep
$p$ as a parameter in the theoretical analysis, and will turn later to
the issue of estimating it.

\subsection{Replica symmetric solution}

It turns out that the ferromagnetically biased Ising spin glass given
by Hamiltonian~(\ref{eq:ham2}) has been studied recently by Castellani
et al. in~\cite{castellani} for fixed connectivity graphs. In the
present context, it is more natural to consider random graphs of
Erd\"os-R\'enyi type, with a Poissonian distribution of connectivity.
However, such a change from fixed to Poissonian connectivity usually
does not induce any qualitative change in the phase diagram.

Castellani et al. use the cavity method~\cite{cavity} to compute,
among other quantities, the one we are interested in: the ground state
magnetization as a function of the parameter $p$. Let us
summarize briefly their main results: at low $p$, the ground state is
replica symmetric and magnetized, the ground state magnetization
approaching $1$ when $p$ goes to $0$; at some critical $p=p_{RSB}$,
the replica symmetry is broken, but the ground state is still
magnetized; finally, for $p>p_c$, the ground state looses its
magnetization. When the connectivity of the graph increases, this
picture is unchanged, but the value of $p_{RSB}$ and $p_c$ increase.

\begin{figure}
\psfrag{k}{$k$}
\psfrag{u1}{$u_1$}
\psfrag{uk}{$u_k$}
\psfrag{u0}{$u$}
\psfrag{k1}{$k_1$}
\psfrag{k2}{$k_2$}
\psfrag{J}{$J$}
\centering{\resizebox{0.7\textwidth}{.16\textheight}{\includegraphics{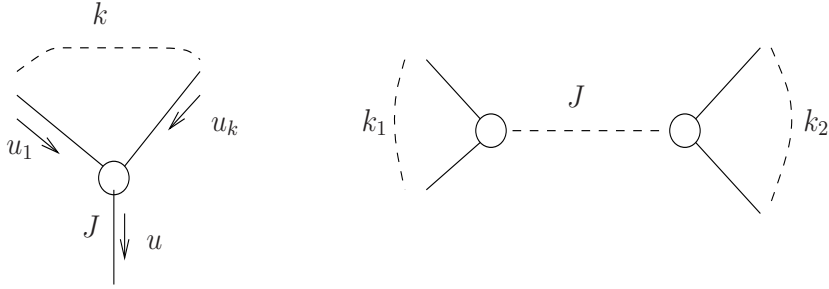}}}
\caption{Left: schematic representation of an iteration, leading to 
Eqs.~(\ref{eq:recursion}). The fields $u_1,\ldots,u_k$ are all equal to $0, 1$ 
or $-1$; $k_0, k_+$ and $k_-$ are respectively the number of fields equal 
to $0, 1$ and $-1$. Right: schematic representation of the addition of a 
link with coupling $J$, leading to Eq.~(\ref{eq:addlink}).
\label{fig:recursion}}
\end{figure}

We give now the replica symmetric solution of~(\ref{eq:ham2}), for an
Erd\"os-R\'enyi random graph, with a Poissonian connectivity distribution 
$\pi(k)$, of degree $\gamma$. 
$$
\pi(k)=e^{-\gamma}\frac{\gamma^k}{k!}~.
$$
The calculations closely follow those
of~\cite{castellani} for fixed connectivity. The cavity messages $u$
sent by the sites along the links take only the values $+1$, $-1$ and
$0$. At the replica symmetric level, the system is then described by a
single probability distribution:
\begin{equation}
{\cal P}(u)=q_+\delta(u-1)+q_0\delta(u)+q_-\delta(u+1)~.
\end{equation}
We write a recursion relation for the probability distribution
${\cal P}$ as follows:
\begin{equation}
{\cal P}(u)=\sum_{k=0}^{\infty}e^{-\gamma}\frac{\gamma^k}{k!}E_J\int
\Pi_{i=1}^{k}d{\cal P}(u_i)~\delta\left(u-\mbox{sgn}(J\sum_{i=1}^k u_i) \right)~,
\label{eq:recursion1}
\end{equation}
where sgn is the sign function, taken to be zero when the argument is
zero; $E_J$ means ``expectation'' over the coupling $J$.
Eq.~\ref{eq:recursion1} straightforwardly translates into three fixed
point equations for $q_0,q_+$ and $q_-$ (see Fig.~\ref{fig:recursion}
for an explanation of $k_+,k_-$ and $k_0$):
\begin{eqnarray}
  q_0&=&\sum_{k=0}^{\infty}e^{-\gamma}\gamma^k\sum_{k_0=0}^k 
\sum_{\stackrel{k_+=k_-}{k_0+k_++k_-=k}}\frac{q_+^{k_+}q_-^{k_-}q_0^{k_0}}{k_+!k_-!k_0!} 
\nonumber \\
  q_+&=&\sum_{k=0}^{\infty}e^{-\gamma}\gamma^k\left[ (1-p)
\sum_{k_0=0}^k \sum_{\stackrel{k_+>k_-}{k_0+k_++k_-=k}}
\frac{q_+^{k_+}q_-^{k_-}q_0^{k_0}} 
{k_+!k_-!k_0!} +p \sum_{k_0=0}^k \sum_{\stackrel{k_+<k_-}{k_0+k_++k_-=k}} 
\frac{q_+^{k_+}q_-^{k_-}q_0^{k_0}}{k_+!k_-!k_0!}\right]
\nonumber \\
  q_-&=&1-q_+-q_0~.
\label{eq:recursion}
\end{eqnarray}
Once $q_0,q_+$ and $q_-$ are known, the ground state magnetization is
given by the expression:
\begin{equation}
m=\sum_{k=0}^{\infty}e^{-\gamma}\gamma^k\left[\sum_{k_0=0}^k 
\sum_{\stackrel{k_+>k_-}{k_0+k_++k_-=k}}\frac{q_+^{k_+}q_-^{k_-}q_0^{k_0}}{k_+!k_-!k_0!}
-\sum_{k_0=0}^k \sum_{\stackrel{k_+<k_-}{k_0+k_++k_-=k}} 
\frac{q_+^{k_+}q_-^{k_-}q_0^{k_0}}{k_+!k_-!k_0!}\right]~.
\end{equation}
To compute the ground state energy, one computes the energy shifts
$\Delta E_s$ due to the addition of a site, and $\Delta E_l$ due to
the addition of a link. One gets after straightforward calculations:
\begin{eqnarray}
\Delta E_s&=&\sum_{k=0}^{\infty}e^{-\gamma}\gamma^k \sum_{k_0+k_++k_-=k}
\frac{q_+^{k_+}q_-^{k_-}q_0^{k_0}}{k_+!k_-!k_0!}(-k_0-|k_+-k_-|)  \\
\Delta E_l&=& -\frac{(q_+-q_-)^2}{1-2p}-2q_0+q_0^2~. 
\label{eq:addlink}
\end{eqnarray}
The ground state energy $e_{gs}$ is then given by
\begin{equation}
e_{gs}=\Delta E_s-\frac{\gamma}{2}\Delta E_l~.
\end{equation}

The qualitative picture emerging from this replica symmetric analysis
is the following: for each mean connectivity $\gamma>1$, there is a
critical value $p_c^{RS}(\gamma)$ such that for $p<p_c^{RS}(\gamma)$,
it is possible to extract information from the knowledge network. The
error rate $\varepsilon$ in the $N\to \infty$ limit is directly
related to the ground state magnetization $m$~:
$$
\varepsilon(p,\gamma)=\frac{1-m(p,\gamma)}{2}~.
$$
For $p>p_c^{RS}(\gamma)$, it is not possible any more to extract
meaningful information from the data in the limit $N\to \infty$: the
error rate tends to $1/2$.

\subsection{Discussion}

We compare these replica symmetric analytic results to numerical
simulations in the next section. We can make however some a priori
remarks on the validity of the calculation. First, we expect the
calculations to be exact at small enough $p$; we then expect a replica
symmetry breaking transition at some
$p_{RSB}(\gamma)<p_c^{RS}(\gamma)$. For $p>p_{RSB}(\gamma)$, the
replica symmetric results are not reliable any more. We expect that
the phase transition described above towards a non magnetized ground
state is shifted to some $p_c^{RSB}\neq p_c^{RS}$. However, the
qualitative result of a transition between one phase which contains some
information and another one which does not should still hold true.

Another word of caution is in order: the authors of~\cite{castellani}
note strong finite size effects for a fixed connectivity network; this
is likely to be the case also for a Poissonian network, and it may
smear out somewhat the transition for finite $N$.

%These results directly translate in our context: for $p<p_{RSB}$, we
%can extract information from our knowledge network, and the
%corresponding optimization problem is ``easy''; for $p_{RSB}<p<p_c$,
%it is still possible to extract information from the knowledge
%network, but the corresponding optimization problem is ``hard'';
%finally, for $p>p_c$, it is not possible any more to extract
%meaningful information from the datas. Regarding this phase transition
%around $p_c$, a word of caution is in order: the authors of
%~\cite{castellani} note strong finite size effects. This means that
%the ground state magnetization tends very slowly to zero for large
%$N$; however, these finite size effects tend to increase the amount of
%information that can be extracted from the network.

%We reproduce here some of their results, for completeness?

\subsection{Estimating $p$}

As already noted above, Eq.~(\ref{eq:ham2}) only depends on $p$
through the parameter $B$, so an a priori knowledge of $p$ is not
necessary to carry out the minimization. This is an interesting practical
advantage.  However, the amount of errors contained in the minimizer
strongly depends on $p$, as explained above. So it would be useful to
have some information about the value of $p$, to get an estimate of
the amount of errors contained in the ground state. It is indeed in
some cases possible to estimate $p$ from the only available data, the
$J_{ij}$'s. Suppose we are given a network. It is possible to compute
for this network $e_{GS}(p)$, the ground state energy as a function of
$p$, by randomly choosing the $J_{ij}$'s with probability $p$; this
can be done analytically in some cases with the cavity method, or
numerically. Then one computes the ground state of the network with
the real $J_{ij}$'s from the data; comparing with the $e_{GS}(p)$, one 
gets an estimate of $p$, provided the $e_{GS}(p)$ curve is not flat.

%indep of p...\\
%estimate of p...\\

\section{Numerical simulations}
\label{sec:num}

We now compare the analytical prediction of the previous section to
data generated randomly: we randomly assign a value $S_i^{(0)}=1$ or
$S_i^{(0)}=-1$ to $N$ spins; we randomly draw a network connecting
these spins, and randomly assign a value $1$ or $-1$ to each link
$J_{ij}$ connecting spins $i$ and $j$, following the rule: 
\begin{eqnarray}
J_{ij}=& S_i^{(0)}S_j^{(0)} & \mbox{with probability}~1-p~, \nonumber \\
J_{ij}=& -S_i^{(0)}S_j^{(0)}& \mbox{with probability}~p. \nonumber
\end{eqnarray}
We then numerically minimize the corresponding Hamiltonian. 
%\subsection{Simulated annealing}
%In a first series of simulations, we used simulated annealing to find
%an approximate ground state of Hamiltonian~(\ref{eq:ham}).
%\subsection{Belief propagation}
For this purpose, we may use simulated annealing. It is simple to
program, but not very fast, and does not perform well in the replica
symmetry broken phase. However, the structure of the problem may
suggest to use another class of algorithm, intensively studied in
different contexts recently (see for instance~\cite{ecc} for a
pedagogical introduction in the context of error correcting codes):
Belief Propagation (BP). BP is not expected to perform better than
simulated annealing in the replica symmetry broken phase, and it may
sometimes fail to converge.  
% (see Fig.~\ref{Fig:compa_sa-bp} for a comparison between BP and
% simulated annealing).
However, it performs overall very well, and is much faster than
simulated annealing, which allows to reach higher $N$: this is crucial
to deal with large data sets.
%\begin{figure}
%\resizebox{0.47\textwidth}{0.3\textheight}{\includegraphics{compa_sa-bp_en2.eps}} \hfill
%\resizebox{0.47\textwidth}{0.3\textheight}{\includegraphics{compa_sa-bp_er.eps}} 
%\caption{Energy (left) and error rate (right) as a function of $p$, 
%for a 3-regular graph. Stars are 
%obtained from simulated annealing ($N=1000$), and diamonds from BP 
%($N=8000$). The solid line 
%is the replica symmetric result. All coincide at low $p$; at high $p$, 
%simulated annealing finds lower energy configurations than BP, but does not 
%perform better in terms of error rate.  }
%\label{Fig:compa_sa-bp} % caption for the whole figure
%\end{figure}

\begin{figure}
\psfrag{p}{\Large p}
\psfrag{ground state energy}{\Large Ground state energy}
\resizebox{0.47\textwidth}{.3\textheight}{\includegraphics{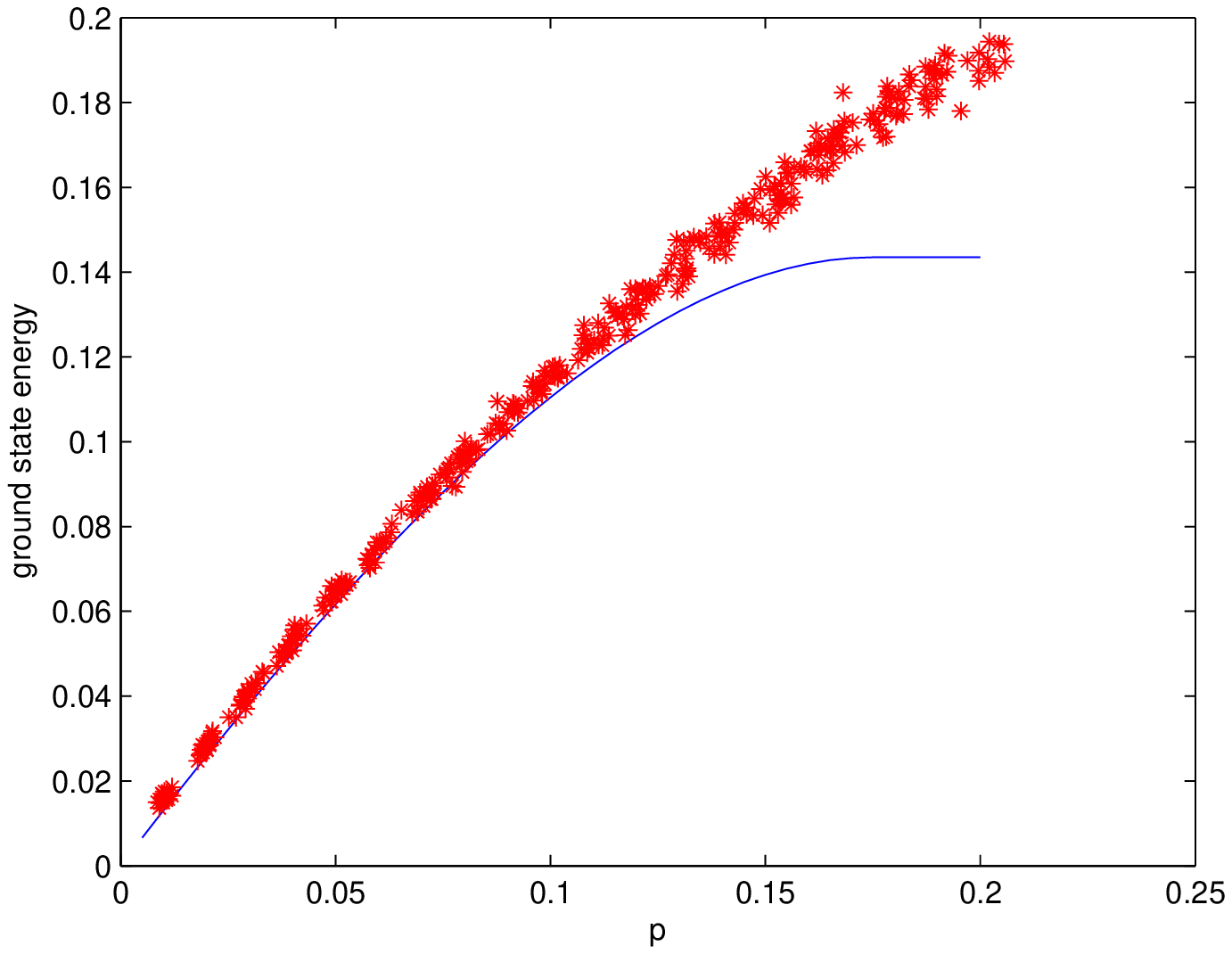}}\hfill
\psfrag{p}{\Large p}
\psfrag{error rate}{\Large Error rate}
\resizebox{0.47\textwidth}{.3\textheight}{\includegraphics{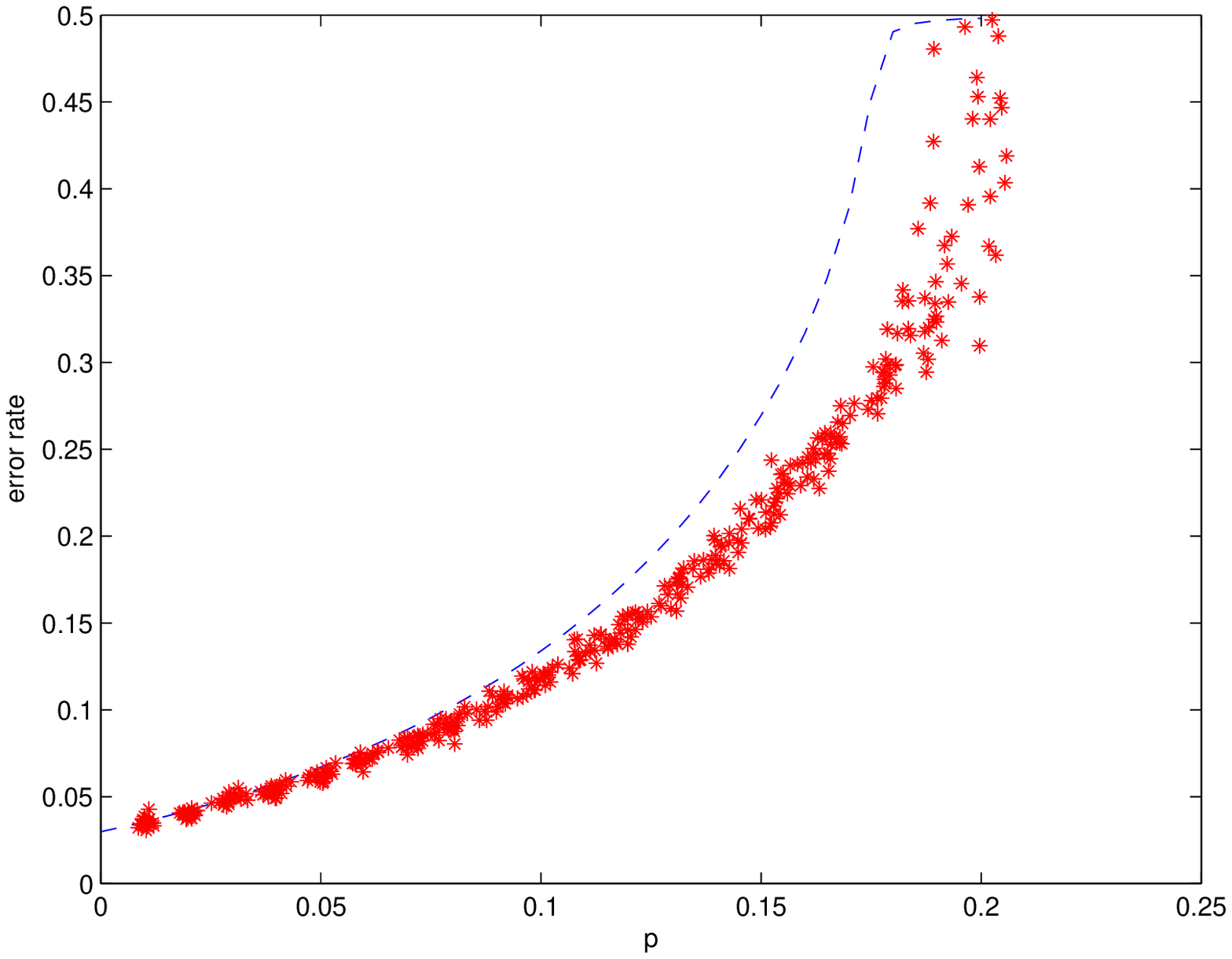}}
%\centering
%\includegraphics[width=6cm]{fig_N8000_g3_en.eps}
\caption{Energy (left) and error rate (right) as a function of $p$, 
for a $\gamma=3$ Poissonian random graph. Symbols are from 
numerical simulations using the BP algorithm, with $N=8000$; the solid 
and dashed lines are the replica symmetric analytical results. 
\label{Fig:energy_magn}}
\end{figure}
%\begin{figure}
%\centering{\resizebox{0.7\textwidth}{.3\textheight}{\includegraphics{fig_N8000_g3_en.eps}}}
%  \caption{Energy as a function of $p$, for $\gamma=3$. Symbols are from 
%numerical simulations using the BP algorithm, with $N=8000$; the solid 
%line is the replica symmetric analytical result. 
%\label{Fig:energy}}
%\end{figure}

%\begin{figure}
%\centering{\resizebox{0.7\textwidth}{.3\textheight}{\includegraphics{fig_N8000_g3.eps}}}
%  \caption{Error rate as a function of $p$, for $\gamma=3$. Symbols are from 
%numerical simulations using the BP algorithm, with $N=8000$; the dashed 
%line is the replica symmetric analytical result. 
%\label{Fig:magn}}
%\end{figure}

On Fig.~\ref{Fig:energy_magn}, one sees that the agreement between
simulations using BP and replica symmetric calculations is very good
for low $p$. For larger $p$, there are important discrepancies, that
may have two origins. First, one expects a replica symmmetry breaking,
as in~\cite{castellani}; this means that the replica symmetric
calculation is not exact any more, and that BP is not expected to
perform well. Second, as already noticed in~\cite{castellani}, finite
size effects are strong. However, the numerical results seem
compatible with the main analytical finding: the presence of a
transition between a low $p$ phase which contains information, and a
high $p$ one that does not. We also note that the error rate obtained
with BP is always smaller than the theoretical one estimated from the
replica symmetric analysis. Finally, it is interesting to compare
quantitatively these results with those of~\cite{castellani} for
regular graphs: both theory and numerics predict a significantly
higher threshold between the informative and non informative phases
for a Poissonian network, for a given mean connectivity.

BP does have another big advantage over simulated annealing: its
outcome is a magnetization for each site; so we also have an
indication on which sites are most likely to be wrongly guessed (those
with magnetization close to zero).  As a final remark, it could be
possible to improve performance in the replica symmetry broken phase
by using a survey propagation algorithm~\cite{Mezard02}.

%BP is much faster than simulated annealing, and allows to reach higher
%$N$; it does not not perfom better than simulated annealing in the
%replica symmetry broken phase however, and it sometimes fails to
%converge~\cite{florent}. It does have another big adavntage over
%simulated annealing: its outcome is a magnetization for each site; so
%we also have an indication on which sites are most likely to be
%wrongly guessed (those with magnetization close to zero).
%To improve performance in this phase, one may think of using a survey 
%propagation algorithm~\cite{Mezard02}.  

%belief propagation...\\

\section{The US Senate example}
\label{sec:US_senate}

\begin{figure}
\psfrag{gamma}{\Large $\gamma$}
\psfrag{error rate}{\Large Error rate}
\centering{\resizebox{0.7\textwidth}{.3\textheight}{\includegraphics{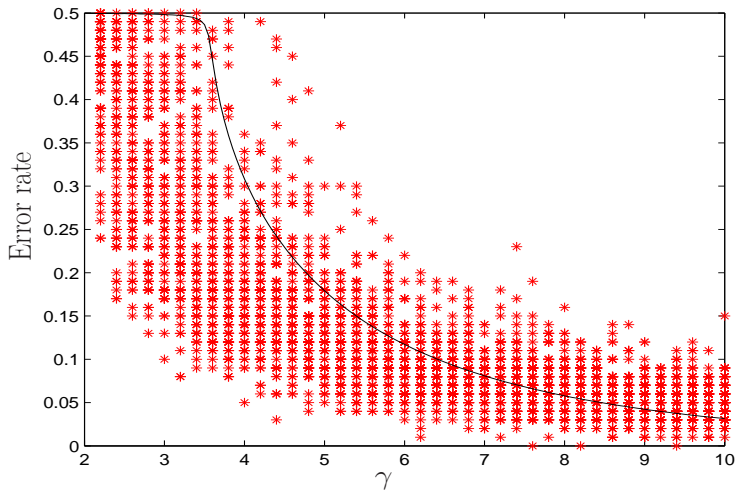}}}
\caption{The parameter $p$ is fixed, $p=0.2$. For each value of
  $\gamma$, the symbols correspond to $100$ realizations.  The solid
  line is the analytical replica symmetric result.
    \label{Fig:USsenate}}
\end{figure}

%\begin{figure}
%\centering{\resizebox{0.7\textwidth}{.3\textheight}{\includegraphics{senators.eps}}}
%  \caption{$p=0.2$, $\gamma=6$. Percentage of errors for each senator, 
%computed for $1000$ runs. The four highest error rates are for senators 
%Miller (M), Inouye (I), Nelson (N) and Specter (S).
%\label{Fig:USsenate2}}
%\end{figure}

The analytical results of section~\ref{sec:cavity} are strengthened by
the numerical simulations of section~\ref{sec:num}; however, unlike
the numerical data, any real data set does not follow exactly the
probabilistic model underlying our study. It is thus important to
assess how robust are the results with respect to some uncertainty in
the model. In this section, we will analyze data from the United
States senate votes, and show that the strategy of minimizing
Hamitonian~(\ref{eq:ham2}) does allow to retrieve some information
from the data; the amount of information retrieved is in
reasonable quantitative agreement with the predictions of
section~\ref{sec:cavity}\footnote{We certainly do not claim that the
  present method is the best possible to extract information from the
  US Senate data; we only try to test the robustness of our results
  on a real data set.}.

%We turn now to an example using real data, which certainly does not
%follow the probabilistic model underlying our study. We will see that
%in spite of this, the strategy of minimizing Hamitonian~(\ref{eq:ham2})
%does allow to retrieve some information from noisy data, and that the
%amount of information retrieved is in reasonable quantitative
%agreement with the predictions of section~\ref{sec:cavity}.

We consider here as agents the 100 US Senators serving in 2001.  The
party of each senator plays the role of the unknown opinion $s_i^0$;
say $s_i^0=-1$ if senator $i$ is a Democrat, and $s_i^0=1$ if senator
$i$ is a Republican. On the US Senate website
(http://www.senate.gov/), the voting positions of all senators are
available for the so-called "roll call votes".
We expect that senators from the same party tend to cast
the same vote, and senators from different parties tend to vote
differently, although it is of course not an absolute rule.

We construct an instance of the "knowledge network problem" as follows:
\begin{itemize}
\item We pick up a random network with given parameter $\gamma$, 
and the senators as nodes. 
\item For each edge of the network, linking two senators with labels
  $i$ and $j$, we pick up randomly one roll call vote in early
  2001\footnote{In practice, we have collected the data from 50 roll
    call votes in early 2001} and consider the voting positions of the
  two senators $i$ and $j$. If they casted the same vote, we set
  $J_{ij}=1$; if they casted a different vote, we set $J_{ij}=-1$.
\end{itemize}
Varying the random network and the random pick of the roll call votes
for each link, we can generate many different instances of the
"knowledge network" for each $\gamma$.

As senators from the same (resp. different) party tend to cast the
same (resp. different) vote, they tend to be linked by edges with
positive (resp. negative) $J$'s. The fact that senators do not always
vote like the majority of their colleagues from the same party plays
the role of a noise.  We crudely model this situation as in
section~\ref{sec:model}, assuming that $J_{ij}=s_i^0s_j^0$ with
probability $1-p$, and $J_{ij}=-s_i^0s_j^0$ with probability $p$, $p$
being unknown, smaller than $1/2$. We now want to retrieve some
information about the $s_i^0$'s (ie the party of each senator), using
the method described in this paper.

Based only on the set of the $J_{ij}$, we run the BP algorithm for
each instance of the "knowledge network", without using any a priori
knowledge on the parameter $p$; we then split the senate in
Republicans and Democrats, according to the BP results. We can check
how many errors we have, and compare with the theory of
section~\ref{sec:cavity}. Note that we can choose the connectivity
of the random network $\gamma$. We have no control however on the
parameter $p$.

The results are presented in Fig.~\ref{Fig:USsenate}, and compared to
the replica symmetric analytical calculations. They seem to be
consistent with the main qualitative analytical result: the existence
of a threshold separating a phase containing almost no information
(low $\gamma$) and a phase which contains some (high $\gamma$).  We
also see on Fig.~\ref{Fig:USsenate}, that there is a strong sample to
sample variability; for small error rates however (large values of the
mean connectivity $\gamma$), the agreement is rather good; for smaller
$\gamma$, the agreement is poor.  There are two explanations for that,
besides the fact that the votes are not random: replica symmetry is
probably broken, and, which is more important for such small systems
($N=100$), finite size effects create large bias. We note however that
the practical error rate is usually smaller than the analytical one.

\section{Conclusion}
\label{sec:conclusion}

We have extended the ``knowledge network'' formalism of~\cite{zhang}
to the more realistic case of noisy data. We have shown that there is
a phase transition between an information-rich phase, and a phase that
essentially contains no information. In the former situation, the
information may be efficiently retrieved through a Belief Propagation
algorithm.

There are several possible extensions to this work. The most direct
ones are the study of non-binary opinions (Potts-like models), or
multidimensionnal opinions. With the applications to commercial
websites in mind presented in~\cite{zhang,franco}, it would also be
interesting to consider bipartite networks. For all these cases, it
seems that the disordered statistical mechanics point of view
used in this paper may be fruitful, by suggesting the use of some
powerful analytical as well as numerical techniques.

%\noindent
\ack
The author thanks Christophe Giraud and Florent Krzakala for discussions
about this work.

\end{document}